\documentclass[prd,preprint,showpacs]{revtex4-1}
\usepackage{amsmath}
\usepackage{latexsym}
\usepackage{amsfonts}
\usepackage{graphicx}
\usepackage{mathrsfs}
\usepackage{hyperref}

\begin{document}

\title{Polarizations of gravitational waves in $f(R)$ gravity}
\author{Dicong Liang}
\email{dcliang@hust.edu.cn}
\affiliation{School of Physics, Huazhong University of Science and Technology,
Wuhan, Hubei 430074, China}
\author{Yungui Gong}
\email{yggong@hust.edu.cn}
\affiliation{School of Physics, Huazhong University of Science and Technology,
Wuhan, Hubei 430074, China}
\author{Shaoqi Hou}
\email{shou1397@hust.edu.cn}
\affiliation{School of Physics, Huazhong University of Science and Technology,
Wuhan, Hubei 430074, China}
\author{Yunqi Liu}
\email{liuyunqi@hust.edu.cn}
\affiliation{School of Physics, Huazhong University of Science and Technology,
Wuhan, Hubei 430074, China}

\begin{abstract}
We point out that there are only three polarizations for gravitational waves in $f(R)$ gravity, and the polarization due to the massive
scalar mode is a mix of the pure longitudinal and transverse breathing polarization.
The classification of the six polarizations by the Newman-Penrose quantities is based on weak, plane and null gravitational waves,
so it is not applicable to the massive mode.
\end{abstract}
\preprint{1701.05998}

\maketitle

\section{Introduction}
The discovery of the gravitational wave event GW150914 by the LIGO Scientific Collaboration
and Virgo Collaboration opens a new window to probe gravitational physics \cite{Abbott:2016blz}.
For example, the detection of GW150914 gave the upper limit of the graviton mass as $m<1.2\times 10^{-22}$ eV \cite{Abbott:2016blz}.
In Einstein's general relativity,
the gravitational waves have two polarizations, the so-called plus and cross modes.
For null plane gravitational waves in general metric theories of gravity, there are six polarization states
denoted by the six independent Newman-Penrose quantities $\Psi_2$, $\Psi_3$, $\Psi_4$ and $\Phi_{22}$
\cite{Newman:1961qr,Eardley:1974nw}.
For Brans-Dicke theory of gravity \cite{Brans:1961sx}, in addition to the plus
and cross modes $\Psi_4$ present in Einstein's gravity, there is another breathing mode $\Phi_{22}$ \cite{Eardley:1974nw}.

The quadratic terms $R_{\mu\nu\alpha\beta} R^{\mu\nu\alpha\beta}$ and $R^2$ are introduced as counterterms
to remove the singularities in the energy-momentum tensor for quantized matter fields interacting with classical gravitational field \cite{Utiyama:1962sn}.
Because there exists the Gauss-Bonnet topological invariance in four dimensions,
\begin{equation}
\label{gaussbonnet1}
\int d^4 x \sqrt{-g}(R_{\mu\nu\alpha\beta} R^{\mu\nu\alpha\beta}-4R_{\mu\nu}R^{\mu\nu}+R^2)=0,
\end{equation}
the general quadratic action can be written as
\begin{equation}
\label{quadractic1}
\int d^4x \sqrt{-g}(aR_{\mu\nu}R^{\mu\nu}+bR^2).
\end{equation}
The addition of the above quadratic terms makes the gravitational theory renormalizable \cite{Stelle:1976gc}.
If $a=0$, then the model $R+\alpha R^2$ introduces an additional massive scalar degree of freedom \cite{Stelle:1977ry}.
In fact, the general nonlinear gravitational theory $f(R)$ is equivalent to a scalar-tensor theory of gravity \cite{OHanlon:1972xqa,Teyssandier:1983zz},
so the $f(R)$ gravity adds an extra massive scalar excitation \cite{Teyssandier:1983zz}.
The polarizations of gravitational waves in $f(R)$ gravity and their detection were discussed in \cite{Corda:2007hi,Corda:2007nr,Capozziello:2008rq,Capozziello:2010iy,Prasia:2014zya}.
It was found that the massive scalar mode in $f(R)$ gravity leads to the longitudinal polarization \cite{Corda:2007hi,Corda:2007nr}.
However, by using the Newman-Penrose formalism \cite{Newman:1961qr,Eardley:1974nw}, the authors claimed that there are four degrees
of freedom in $f(R)$ gravity because the nonzero Newman-Penrose quantities are $\Psi_2$, $\Psi_4$ and $\Phi_{22}$ \cite{Alves:2009eg,Alves:2010ms}.
The point was further explained by arguing that the traceless condition
cannot be imposed \cite{Rizwana:2016qdq}. Myung pointed out that there is no problem for imposing the transverse traceless condition \cite{Corda:2007nr}
and there are only three degrees of freedom in $f(R)$  gravity \cite{Myung:2016zdl}.

The $f(R)$ model with $R+\alpha R^2$ was first applied to cosmology by Starobinsky as an inflationary model \cite{starobinskyfr},
and its predictions are consistent with current observations \cite{Ade:2015lrj}. The $f(R)$ gravity was also invoked
to explain the late time cosmic acceleration discovered by the supernova observations \cite{hzsst98,scpsn98}.
The first such model was $f(R)=R+\alpha R^{-1}$ \cite{Carroll:2003wy,Vollick:2003aw,Flanagan:2003rb,Nojiri:2003ft}, but it was ruled out
by the solar system tests \cite{Chiba:2003ir,Erickcek:2006vf}, so more viable $f(R)$ models were then proposed \cite{Hu:2007nk,Starobinsky:2007hu,Capozziello:2008fn,Cognola:2007zu,Nojiri:2007cq,Myrzakulov:2015qaa,Yi:2016jqr}.

In this paper, we study the polarizations of gravitational waves in $f(R)$ gravity with several different methods. In Sec. II,
we discuss the transverse and traceless condition, the energy current carried by the gravitational waves,
the wave equations for weak gravitational fields around the flat
background, and the particle contents in $f(R)$ gravity.
In Sec. III, we discuss the polarizations from the equivalent scalar-tensor theory with the help of the Newman-Penrose formalism,
and analyze the dynamical degrees of freedom by using the Hamiltonian method for general $f(R)$ theory of gravity.
We point out that the six polarizations of gravitational waves were derived for weak, plane and {\em null} gravitational waves
and the result is not applicable to massive modes.
We conclude the paper in Sec. IV.

\section{Nonlinear Gravity Theory}
\label{1.ngt}

The action is
\begin{equation}
\label{fr}
S={1\over 16\pi G}\int d^4 x\sqrt{-g}\,f(R),
\end{equation}
and the field equation is
\begin{equation}
\label{freinseq1}
f'(R)R_{\mu\nu}-\frac{1}{2}f(R)g_{\mu\nu}-\nabla_\mu\nabla_\nu f'(R)+g_{\mu\nu}\Box f'(R)=0,
\end{equation}
where $\Box=g^{\mu\nu}\nabla_\mu\nabla_\nu$. Taking the trace of Eq. \eqref{freinseq1}, we get
\begin{equation}
\label{freinseq2}
f'(R)R+3\Box f'(R)-2f(R)=0.
\end{equation}
Perturbing the metric around the Minkowski metric $g_{\mu\nu}=\eta_{\mu\nu}+h_{\mu\nu}$, to the first order of $h_{\mu\nu}$, we get
\begin{equation}
\label{frperteq1}
R_{\mu\nu}=\frac{1}{2}\left(\partial_\mu\partial_\rho h^\rho_\nu+\partial_\nu\partial_\rho h^\rho_\mu-\partial_\mu\partial_\nu h-\Box h_{\mu\nu}\right),
\end{equation}
\begin{equation}
\label{frperteq2}
R=\partial_\mu\partial_\rho h^{\rho\mu}-\Box h,
\end{equation}
where $h=\eta^{\mu\nu}h_{\mu\nu}$ and the d'Alembert operator becomes $\Box=\eta^{\mu\nu}\partial_\mu\partial_\nu$.
For the particular model $f(R)=R+\alpha R^2$, to the first order of perturbation, Eq. \eqref{freinseq1} becomes
\begin{equation}
\label{frperteq3}
R_{\mu\nu}-\frac{1}{2}\eta_{\mu\nu}R-2\alpha\left(\partial_\mu\partial_\nu R-\eta_{\mu\nu}\Box R\right)=0.
\end{equation}
Taking the trace of Eq. \eqref{frperteq3} or using Eq. \eqref{freinseq2}, we get
\begin{equation}
\label{frperteq4}
(\Box-m^2)R=0,
\end{equation}
where $m^2=1/(6\alpha)$ with $\alpha>0$. The upper limit of the graviton mass given by the LIGO observations is $m<1.2\times 10^{-22}$ eV \cite{Abbott:2016blz},
and a more stringent limit from the dynamics of the galaxy cluster is $m<2\times 10^{-29}$ eV \cite{Goldhaber:1974wg}.

Introduce the variable
\begin{equation}
\label{canhtt}
\bar{h}_{\mu\nu}=h_{\mu\nu}-\frac{1}{2}\eta_{\mu\nu}h-2\alpha\eta_{\mu\nu} R,
\end{equation}
so
\begin{equation}
\label{canhtt1}
\bar{h}=\eta^{\mu\nu}\bar{h}_{\mu\nu}=-h-8\alpha R,
\end{equation}
\begin{equation}
\label{canhtt2}
h_{\mu\nu}=\bar{h}_{\mu\nu}-\frac{1}{2}\eta_{\mu\nu}\bar{h}-2\alpha\eta_{\mu\nu} R.
\end{equation}
Under an infinitesimal coordinate transformation, $x^\mu\rightarrow x^{\mu\prime}=x^\mu+\epsilon^\mu$, we have
\begin{equation}
\label{coortranfeq21}
h_{\mu\nu}'=h_{\mu\nu}-\partial_\mu\epsilon_\nu-\partial_\nu\epsilon_\mu,
\end{equation}
\begin{equation}
\label{coortranfeq22}
h'=h-2\partial_\mu\epsilon^\mu,
\end{equation}
\begin{equation}
\label{coortranfeq23}
\bar{h}_{\mu\nu}'=\bar{h}_{\mu\nu}-\partial_\mu\epsilon_\nu-\partial_\nu\epsilon_\mu+\eta_{\mu\nu}\partial_\rho\epsilon^\rho,
\end{equation}
\begin{equation}
\label{coortranfeq24}
\bar{h}'=\bar{h}+2\partial_\rho\epsilon^\rho,
\end{equation}
where the index was raised or lowered by the Minkowski metric $\eta_{\mu\nu}$, i.e., $\epsilon_\mu=\eta_{\mu\nu}\epsilon^\nu$.
If we choose $\epsilon_\mu$ so that it satisfies the equation
\begin{equation}
\label{coortranfeq25}
\Box\epsilon_\nu=\partial^\mu\bar{h}_{\mu\nu},
\end{equation}
then we get the Lorenz gauge condition $\partial^\mu \bar{h}_{\mu\nu}'=0$. 
Note that the Lorenz condition does not fix the gauge freedom completely;
it leaves a residual coordinate transformation $x^{\mu\prime}=x^\mu+\xi^\mu$ with $\Box\xi^\mu=0$. If $\xi^\mu$ also satisfies the equation
$\partial_\mu\xi^\mu=-\bar{h}/2$, then we get $\bar{h}'=0$. Therefore,
it is always possible to choose the transverse traceless gauge condition \cite{Corda:2007hi,Corda:2007nr,Capozziello:2008rq}
\begin{equation}
\label{gaugeeq1}
\partial^\mu \bar{h}_{\mu\nu}=0,\quad \bar{h}=\eta^{\mu\nu}\bar{h}_{\mu\nu}=0.
\end{equation}
By using the transverse traceless gauge condition, substituting Eq. \eqref{canhtt2} into Eq. \eqref{frperteq1}, we get
\begin{equation}
\label{frperteq6}
R_{\mu\nu}=\frac{1}{2}\left[-\Box\bar{h}_{\mu\nu}+4\alpha\partial_\mu\partial_\nu R+2\alpha\eta_{\mu\nu}\Box R\right].
\end{equation}
The trace of Eq. \eqref{frperteq6} gives Eq. \eqref{frperteq4}. Plugging Eq. \eqref{frperteq6} into Eq. \eqref{frperteq3}, we get
\begin{equation}
\label{frperteq7}
3\alpha\eta_{\mu\nu}(\Box-m^2)R-\frac{1}{2}\Box\bar{h}_{\mu\nu}=0.
\end{equation}
Combining Eqs. \eqref{frperteq7} and \eqref{frperteq4}, we get
\begin{equation}
\label{frperteq8}
\Box\bar{h}_{\mu\nu}=0.
\end{equation}
The solution to Eq. \eqref{frperteq8} is
\begin{equation}
\label{freq29a}
\bar{h}_{\mu\nu}=e_{\mu\nu}\exp(iq_\mu x^\mu)+\text{c.c.},
\end{equation}
where $\eta_{\mu\nu}q^\mu q^\nu=0$ and $q^\mu e_{\mu\nu}=0$.

For null gravitational waves traveling along
the $z$ direction with $q^\mu=\omega(1,0,0,1)$, the energy current is \cite{Berry:2011pb}
\begin{equation}
\label{freq9}
t_{0z}=\frac{1}{8\pi G}\langle G^{(1)}_{0z}-G_{0z}\rangle=\frac{1}{16\pi G}\left\langle
\omega^2\left[\left(\frac{e_{xx}-e_{yy}}{2}\right)^2+e_{xy}^2\right]+48\alpha^2 R_{,0}R_{,z}\right\rangle,
\end{equation}
where $G^{(1)}_{\mu\nu}$ is the first order Einstein tensor.
In deriving the above result, we used the solution \eqref{freq29a} and the transverse condition $q^\mu e_{\mu\nu}=0$,
but we do not apply the traceless condition. From Eq. \eqref{freq9}, it is clear that
a null wave for which $e_{xx}-e_{yy}$ and $e_{xy}$ vanish does not transport energy,
i.e., a null wave with nonzero trace $\bar{h}$ in which $e_{xx}+e_{yy}\neq 0$ does not transport energy, so
the trace $\bar{h}$ is not a physical degree of freedom and
the physical plane wave $\bar{h}_{\mu\nu}$ is transverse and traceless.

From Eqs. \eqref{frperteq4} and \eqref{frperteq8}, we see that under the transverse traceless gauge condition \eqref{gaugeeq1},
the model $f(R)=R+\alpha R^2$ has two massless tensor and one massive scalar degrees of freedom.
This point is also clear from the action. To the second order of $h_{\mu\nu}$, the action becomes \cite{Stelle:1977ry}
\begin{equation}
\label{weak}
\begin{split}
S&=\frac{1}{16\pi G}\int d^4 x \left[{1\over 4}h^{\mu\nu}\Box h_{\mu\nu}-{1\over 2}
h^{\mu\nu}\partial_\mu\partial^\rho h_{\rho\nu}-{1\over 4}h\Box h+
{1\over 4}h\partial_\mu\partial_\nu h^{\mu\nu}+
{1\over 4}h^{\mu\nu}\partial_\mu\partial_\nu h  \right.\\
&\left. \vphantom{{1\over 4}} +\alpha
h^{\mu\nu}\partial_\mu\partial_\nu\partial_\alpha\partial_\beta
h^{\alpha\beta} +\alpha h\Box^2 h -\alpha
h^{\mu\nu}\partial_\mu\partial_\nu\Box h-
\alpha h\Box \partial_\mu\partial_\nu h^{\mu\nu}\right] \\
&=\int d^4 x \frac{1}{32\pi G}h_{\mu\nu}\,P^{\mu\nu,\alpha\beta}
h_{\alpha\beta},
\end{split}
\end{equation}
where
\begin{equation}\label{spineq1}
P^{\mu\nu,\alpha\beta}=\left[{1\over 2}P^{(2)\mu\nu,\alpha\beta}
-P^{(s)\mu\nu,\alpha\beta}\right]\Box
+6\alpha\,P^{(s)\mu\nu,\alpha\beta}\,\Box^2,
\end{equation}
the spin-2 and spin-0
projection operators \cite{spin2}
\begin{equation}
\label{spineq2}
\begin{split}
P^{(2)}_{\mu\nu,\alpha\beta}&=\frac{1}{2}(\theta_{\mu\alpha}\theta_{\nu\beta}+\theta_{\mu\beta}\theta_{\nu\alpha})
- \frac{1}{3}\theta_{\mu\nu}\theta_{\alpha\beta},\\
P^{(s)}_{\mu\nu,\alpha\beta}&={1\over 3}\theta_{\mu\nu}\theta_{\alpha\beta},\\
\theta_{\mu\nu}&=\eta_{\mu\nu}-{\partial_\mu\partial_\nu\over \Box}.
\end{split}
\end{equation}
If $\alpha=0$, then the spin-0 projector $P^{(s)}_{\mu\nu,\alpha\beta}$ is absent and only the spin-2 projector
$P^{(2)}_{\mu\nu,\alpha\beta}$ remains, so Eqs. \eqref{weak} and \eqref{spineq1} reduce to the standard result for the massless spin-2 field.

Because of the diffeomorphism invariance of the theory, the
differential operator $P$ cannot be inverted. In general, we need
to add some gauge fixing terms to the theory so that we can get
the propagator for the spin-2 massless gravitons. Formally, the operator $P$ can be
inverted to give (symbolically, in coordinate space) the quantum
propagator. Therefore, formally we have \cite{Alonso:1994tr}
\begin{equation}
\label{spineq3}
D(h)=-{1\over \Box}(2\,P^{(2)}-P^{(s)})-\frac{1}{\Box-m^2}P^{(s)},
\end{equation}
where $m^2=1/6\alpha$. The first term on the right-hand side of Eq. \eqref{spineq3} denotes the propagator of a massless spin-2 field (graviton) and
the second term denotes the propagator of a massive scalar field. Therefore, the propagating degrees of freedom in $R+\alpha R^2$ gravity are the massless spin-2
gravitons and the massive scalar field with the mass $m^2=1/6\alpha$.

\section{Scalar-tensor theory of gravity}
The action \eqref{fr} can be written as
\begin{equation}
\label{freq21}
S=\frac{1}{2\kappa^2}\int d^4
x\sqrt{-g}\,[f(\varphi)+(R-\varphi)f'(\varphi)]=\frac{1}{2\kappa^2}\int d^4
x\sqrt{-g}\,[f'(\varphi) R+f(\varphi)-\varphi f'(\varphi)],
\end{equation}
where $f'(\varphi)=df(\varphi)/d\varphi$, $\kappa^2=8\pi G$,
and $d^2f(R)/dR^2\neq 0$, so the $f(R)$ gravity is equivalent to the scalar-tensor theory of gravity \cite{OHanlon:1972xqa, Teyssandier:1983zz}.

\subsection{Hamiltonian analysis}

The Hamiltonian formulation was derived before for different $f(R)$ models \cite{Ezawa:1998ax,Ezawa:2005zr,Ohkuwa:2014mwa,Olmo:2011fh,Deruelle:2009pu,Deruelle:2009zk,Sendouda:2011hq}.
In this section, we perform the Hamiltonian analysis for the $f(R)$ gravity with the action (\ref{freq21}) to derive the dynamical degrees of freedom of the theory.
For convenience, we use the Arnowitt-Deser-Misner (ADM) foliation \cite{adm1,adm2} of the spacetime, so the metric is written as
\begin{equation}
\label{admf}
d{s^2} =  - {N^2}d{t^2} + {h_{ij}}\left( {d{x^i} + {N^i}dt} \right)\left( {d{x^j} + {N^j}dt} \right),
\end{equation}
where $N$, $N^i$, $h_{ij}$ are the lapse function, the shift function and the metric for the three-dimensional space, respectively.

Let $n_\mu$ be the unit normal to the constant time slice $\Sigma_t$, so that $n_\mu=-N\nabla_\mu t$, and the exterior curvature of $\Sigma_t$ is
\begin{equation}\label{exK}
  K_{\mu\nu}=\nabla_\mu n_\nu+n_\mu n^\rho\nabla_\rho n_\nu,
\end{equation}
with the spatial components
\begin{equation}\label{exKs}
  K_{jl}=\frac{1}{2N}(\dot h_{jl}-2D_{(j}N_{l)}),
\end{equation}
where $D_j$ represents the covariant derivative with respect to the three-dimensional metric $h_{jl}$,
and the brackets in the subscript imply symmetrization. In terms of ADM variables, the action (\ref{freq21}) becomes
\begin{equation}\label{admact}
  S=\frac{1}{2}\int d^4xN\sqrt{h}\{f+f'[\mathscr R+K_{jl}K^{jl}-K^2+2\nabla_\mu(n^\mu\nabla_\nu n^\nu)-2\nabla_\mu(n^\nu\nabla_\nu n^\mu)-\varphi]\},
\end{equation}
where $\kappa$ is set to 1 for simplicity, $\mathscr R$ is the Ricci tensor for the spatial metric $h_{jl}$ and $K=h^{jl}K_{jl}$ is the trace of $K_{jl}$.
Integration by parts brings the action \eqref{admact} to the following form:
\begin{eqnarray}
  S &=& \int d^4xN\sqrt{h}\Big[\frac{1}{2}f'(\mathscr R-\varphi)+\frac{1}{2}f+\frac{1}{2}f'(K_{jl}K^{jl}-K^2)
  \nonumber\\
  &&+\frac{K}{N}(N_jD^jf'-f''\dot\varphi)+D_j f' D^j\ln N\Big].
\end{eqnarray}
In this action, we have 11 dynamical variables $N$, $N_i$, $h_{ij}$ and $\varphi$. The corresponding canonical momenta are
\begin{eqnarray}
  \pi^N &=& \frac{\delta S}{\delta \dot N}=0,\\
  \pi^j&=&\frac{\delta S}{\delta \dot N_j}=0, \\
  \pi^{jl} &=& \frac{\delta S}{\delta \dot h_{jl}}=\frac{\sqrt{h}}{2}\Big[f'(K^{jl}-h^{jl}K)+\frac{h^{jl}}{N}(N_kD^kf'-f''\dot\varphi)\Big], \\
  p &=& \frac{\delta S}{\delta \dot\varphi}=-\sqrt{h}f''K.
\end{eqnarray}
Therefore, $\pi^N\approx0$ and $\pi^j\approx0$ constitute the primary constraints. These equations can be inverted to solve for $\dot h_{jl}$ and $\dot\varphi$,
\begin{eqnarray}
  \dot h_{jl} &=& \frac{4N}{\sqrt{h}f'}\Big[\pi_{jl}-\frac{1}{3}h_{jl}\Big(\pi+\frac{1}{2}p\frac{f'}{f''}\Big)\Big],\label{dh}\\
  \dot\varphi &=& N_jD^j\varphi+\frac{2N}{3\sqrt{h}f''}\Big(p\frac{f'}{f''}-\pi\Big), \label{dphi}
\end{eqnarray}
in which $\pi=h_{jl}\pi^{jl}$.

The Legendre transformation leads to the following Hamiltonian,
\begin{eqnarray}
  H&=&\int_{\Sigma_t}d^3x(\pi^{jl}\dot h_{jl}+p\dot\varphi-\mathscr L)
  \nonumber\\
  &=&\int_{\Sigma_t}d^3x\sqrt{h}(NC+N_jC^j),
\end{eqnarray}
where the boundary terms have been dropped. The Hamiltonian constraint $C$ and the momentum constraint $C^j$ are
\begin{eqnarray}
    C&=&-\frac{1}{2}[f+f'(\mathscr R-\varphi)]+D_jD^jf'+\frac{2}{hf'}\left(\pi^{jl}\pi_{jl}-\frac{\pi^2}{3}\right)
    \nonumber\\
    &&-\frac{2}{3hf''}\pi p+\frac{1}{3hf'}\left(p\frac{f'}{f''}\right)^2, \label{hc}\\
    C^j&=&\frac{p}{\sqrt{h}}D^j\varphi-2D_l\frac{\pi^{jl}}{\sqrt{h}}.\label{mc}
\end{eqnarray}
These are the secondary constraints, so we have a total of eight constraints.
To obtain the constraint algebra, we choose arbitrary functions $\nu$ and $\mu$,
and arbitrary spatial vectors $v^j$ and $u^j$, all defined on $\Sigma_t$, to define smeared quantities in the following way:
\begin{eqnarray}
\pi_\nu&=&\int_{\Sigma_t}d^3x\sqrt{h}\nu\pi^N,\\
\pi_{\vec v}&=&\int_{\Sigma_t}d^3x\sqrt{h}v_j\pi^j,\\
  C_\mu &=& \int_{\Sigma_t}d^3x \sqrt{h}\mu C, \\
  C_{\vec u} &=& \int_{\Sigma_t} d^3x\sqrt{h}u_jC^j.
\end{eqnarray}
The primary constraints have vanishing Poisson brackets with each other and with $C_\nu$ and $C_{\vec v}$.
The remaining constraint algebra turns out to be
\begin{eqnarray}
  \{C_\nu,C_\mu\} &=& C_{\vec \xi}, \label{hhpb}\\
  \{C_\nu,C_{\vec u}\} &=& -C_{\mathscr L_{\vec{u}}\nu}, \label{hmpb}\\
   \{C_{\vec v},C_{\vec u}\}&=& C_{\vec \zeta},\label{mmpb}
\end{eqnarray}
where the spatial vectors $\vec{\xi}$ and $\vec{\zeta}$ are
\begin{equation}
  \xi^j=\nu D^j\mu-\mu D^j\nu,\quad \zeta^j=u^kD^jv_k-v^kD^ju_k,
\end{equation}
and $\mathscr L_{\vec u}\nu=u^jD_j\nu$ is the three-dimensional Lie derivative.
Therefore, all of the above eight constraints $\pi_\nu$, $C_\mu$, $\pi_{\vec{v}}$ and $C_{\vec{u}}$ are first class.
Since $H=C_N+C_{\vec N}$, the consistency conditions are naturally satisfied,
\begin{eqnarray}
  \{C_\nu,H\} &=& 0, \\
  \{C_{\vec v},H\} &=& 0.
\end{eqnarray}
So there are no further secondary constraints. In the phase space, 
we have 22 dynamical variables and eight first class constraints, so the number of
degrees of freedom for $f(R)$ gravity is $n=(22-8\times 2)/2=3$.

In summary, as in Einstein's general relativity, the vanishing of $\pi^N$ and $\pi^j$ renders $N$ and $N_j$ as Lagrangian multipliers, and thus,
they cease to be dynamical variables. The Hamiltonian and momentum constraints remove four more degrees of freedom of the theory,
and finally, a suitable choice of coordinate conditions further removes four degrees of freedom,
reducing the dimension of the phase space of $f(R)$ gravity to 6. Therefore,
the configuration space of $f(R)$ gravity is three dimensional.

\subsection{The polarization states}

The field equations to the action \eqref{freq21} are
\begin{equation}
\label{freq22}
G_{\mu\nu}=\frac{1}{f'(\varphi)}\left[\nabla_\mu\nabla_\nu f'(\varphi)-g_{\mu\nu}\Box f'(\varphi)+\frac{1}{2}g_{\mu\nu}[f(\varphi)-\varphi f'(\varphi)]\right],
\end{equation}
\begin{equation}
\label{freq23}
\Box f'=\frac{2}{3}f(\varphi)-\frac{1}{3}\varphi f'(\varphi).
\end{equation}
For the model $f(R)=R+\alpha\,R^2$, Eqs. \eqref{freq22} and \eqref{freq23} become
\begin{equation}
\label{freq24}
G_{\mu\nu}=\frac{2\alpha}{1+2\alpha\varphi}\left(\nabla_\mu\nabla_\nu\varphi-g_{\mu\nu}\Box\varphi-\frac{1}{4} g_{\mu\nu}\varphi^2\right),
\end{equation}
\begin{equation}
\label{freq25}
(\Box-m^2)\varphi=0.
\end{equation}
From Eqs. \eqref{freq24} and \eqref{freq25}, we see that the massive scalar field $\varphi$ is the source of the massless spin-2 gravitational field, so there
are one massive scalar mode and two massless tensor modes.
In terms of the variable $\bar{h}_{\mu\nu}=h_{\mu\nu}-\eta_{\mu\nu}h/2-2\alpha\eta_{\mu\nu}\delta\varphi$,
and under the Lorenz gauge condition $\partial^\mu \bar{h}_{\mu\nu}=0$, to the first order
of perturbation around the flat spacetime, we get $G_{\mu\nu}=-\Box\bar{h}_{\mu\nu}/2+2\alpha\partial_\mu\partial_\nu\delta\varphi-2\alpha\eta_{\mu\nu}\Box\delta\varphi$.
Comparing with Eq. \eqref{freq24}, we obtain
\begin{equation}
\label{freq26}
\Box\bar{h}_{\mu\nu}=0,
\end{equation}
\begin{equation}
\label{freq27}
(\Box-m^2)\delta\varphi=0.
\end{equation}
The solutions to the wave equations \eqref{freq26} and \eqref{freq27} are Eq. \eqref{freq29a} and
\begin{equation}
\label{freq29}
\delta\varphi=\phi_1 \exp(ip_\mu x^\mu)+\text{c.c.},
\end{equation}
where $\eta_{\mu\nu}p^\mu p^\nu=-m^2$.
As discussed in the previous section, $\bar{h}_{\mu\nu}$ is transverse and traceless and it denotes the standard spin-2 graviton.
For the plane wave traveling along the $z$ direction, we have $q^\mu=\omega(1,0,0,1)$ and $p^\mu=(\Omega,0,0,\sqrt{\Omega^2-m^2})$.
The speed of the massless spin-2 graviton $\bar{h}_{\mu\nu}$ is the light speed $c=1$ and the speed of the massive scalar field $\delta\varphi$ is $v=\sqrt{\Omega^2-m^2}/\Omega$,
so $h_{\mu\nu}$ is the combination of the function of $t-z$ and the function of $vt-z$,
\begin{equation}
\label{freq31}
h_{\mu\nu}=\bar{h}_{\mu\nu}(t-z)-2\alpha\eta_{\mu\nu}\delta\varphi(vt-z).
\end{equation}

Plugging the solution \eqref{freq29} into Eq. \eqref{freq24}, we get
\begin{equation}
\label{freq30}
R_{\mu\nu}\approx \frac{1}{6}\eta_{\mu\nu}\delta\varphi-2\alpha p_\mu p_\nu\delta\varphi,
\end{equation}
so the nonzero components of the Ricci tensor are $R_{tt}$, $R_{tz}$ and $R_{zz}$ for waves traveling along the $z$ direction.
If the scalar field is massless (like the scalar field in Brans-Dicke theory),
then we can apply the classification based on the Newman-Penrose formalism \cite{Newman:1961qr,Eardley:1974nw} to obtain the effect of the scalar field on the geometry.
In Brans-Dicke theory, the massless scalar field manifests itself as the breathing mode with $\Phi_{22}=-R_{xtxt}-R_{ytyt}\neq 0$ \cite{Eardley:1974nw}.

If we apply the Newman-Penrose formalism, we may get $\Psi_2=-R_{kl}/6\neq 0$, $\Psi_4\neq 0$ and $\Phi_{22}=-R_{ll}/2\neq 0$.
However, we also get $R_{kk}=-2\alpha (p_\mu k^\mu)^2\delta\varphi \neq 0$ which is inconsistent with the Newman-Penrose result
in Ref. \cite{Eardley:1974nw} (see Appendix A for details).
The inconsistency arises because the result based on the Newman-Penrose formalism is derived for waves at the speed of light $c$.
For $f(R)$ gravity, we have a massive scalar
field whose speed is not $c$, so we cannot directly apply the Newman-Penrose formalism to claim that there are more than three polarization modes
based on the result that $\Psi_2\neq 0$.

To understand the polarization state of the massive scalar field, we study its effect on the geodesic deviation.
To the linear order, we get
\begin{equation}
\label{freq32}
R_{\mu\nu\alpha\beta}\approx \frac{1}{2}\left(h_{\nu\alpha,\mu\beta}+h_{\mu\beta,\nu\alpha}-h_{\mu\alpha,\nu\beta}-h_{\nu\beta,\mu\alpha}\right).
\end{equation}
For the massive scalar mode, we have
\begin{equation}
\label{freq33}
R_{itjt}=-\alpha(\delta_{ij}\delta\ddot\varphi-\delta\varphi_{,ij}),
\end{equation}
and the geodesic deviation due to the massive scalar mode is
\begin{equation}
\label{freq34}
\ddot{x}=\alpha \delta\ddot\varphi\, x,
\end{equation}
\begin{equation}
\label{freq35}
\ddot{y}=\alpha \delta\ddot\varphi\, y,
\end{equation}
\begin{equation}
\label{freq36}
\ddot{z}=-\alpha m^2 \delta\varphi\, z=-\frac{1}{6}\delta\varphi\, z.
\end{equation}
Therefore, the polarization of the massive scalar mode for the model $f(R)=R+\alpha R^2$ is a mix of the pure longitudinal and the breathing mode.
Note that the longitude mode is independent of the mass parameter $\alpha$ and there is no massless limit for finite $\alpha$ in the model considered here.
If $\alpha=0$, then $\delta\varphi=0$ and we recover the standard massless plus and cross polarizations.

The propagating speed of the massive mode is less than the speed of the massless mode, so the gravitational wave due to
the massive mode arrives at the detector later. The mix of transverse breathing mode and the longitudinal mode for the polarization
state is a distinct character of the massive mode, so the detection of this polarization state by the network of
the Advanced Laser Interferometer Gravitational-Wave Observatory and Virgo detectors or
the Laser Interferometer Space Antenna can be used to test different gravitational theories.

\section{Conclusions}
We derived the linear wave equations for the model $f(R)=R+\alpha R^2$ and performed the
Hamiltonian analysis for the general $f(R)$ gravity; we found that the propagating degrees of freedom
in $f(R)$ gravity are the two massless spin-2 modes and one massive scalar mode. With
the coordinate transformation, we showed explicitly that the transverse and traceless gauge conditions
can be imposed on the perturbation $\bar{h}_{\mu\nu}$. It was also shown that
the gravitational waves $\bar{h}_{\mu\nu}$ traveling along the $z$ direction for which $\bar{h}_{xx}-\bar{h}_{yy}$ and $\bar{h}_{xy}$
vanish do not transport energy, so the physical $\bar{h}_{\mu\nu}$ must be transverse and traceless.
Working with the equivalent scalar-tensor theory of gravity for the $f(R)$ gravity, we get
a massive scalar field in addition to the massless tensor field $\bar{h}_{\mu\nu}$.
If we apply the Newman-Penrose formalism to gravitational waves with a massive mode, then
we get the longitudinal mode with $\Psi_2\neq 0$ and the breathing mode with $\Phi_{22}\neq 0$
in addition to the plus and cross modes with $\Psi_4\neq 0$, and we also get nonzero $R_{kk}$,
which should be 0 in the Newman-Penrose formalism. The reason for the inconsistency is because
the classification of the polarizations based on the Newman-Penrose formalism is derived for
the weak, plane and null waves. When the massive scalar mode appears, the
classification based on the Newman-Penrsoe formalism is not applicable.
By working out the geodesic deviation for the massive scalar mode, we find that the polarization
is a mix of the longitudinal and transverse breathing mode. For null gravitational waves,
the longitudinal and the breathing modes are independent of each other and they are two different degrees of freedom.
However, for the massive mode, the polarization is a mix of the two and has only one state.
In conclusion, there are only three propagating degrees of freedom in $f(R)$ gravity; two of them
are the massless plus and cross polarizations, and the other massive scalar mode is a mix of longitudinal and transverse polarization.
The potential detection of the massive mode by the ground or space interferometer detectors can be used to distinguish different gravitational theories.

\begin{acknowledgements}
Y. G. thanks the center for quantum spacetime in Sogang University for the hospitality.
This research was supported in part by the National Natural Science
Foundation of China under Grant No. 11475065 and
the Major Program of the National Natural Science Foundation of China under Grant No. 11690021.
\end{acknowledgements}

\appendix
\section{Newman-Penrose formalism}
In the Newman-Penrose formalism \cite{Newman:1961qr}, we introduce the following tetrad system of null vectors \cite{Eardley:1974nw},
\begin{equation}
\label{nptetrad}
\begin{split}
k^\mu=\frac{1}{\sqrt{2}}(e^\mu_t+e^\mu_z),\quad l^\mu=\frac{1}{\sqrt{2}}(e^\mu_t-e^\mu_z),\\
m^\mu=\frac{1}{\sqrt{2}}(e^\mu_x+ie^\mu_y),\quad \bar{m}^\mu=\frac{1}{\sqrt{2}}(e^\mu_x-ie^\mu_y),\\
-k^\mu l_\mu=m^\mu\bar{m}_\mu=1,\quad E^\mu_a=(k^\mu,l^\mu,m^\mu,\bar{m}^\mu),
\end{split}
\end{equation}
where the tetrad indices ($a,b,c,\ldots$) range over $(1,2,3,4)=(k,l,m,\bar{m})$ and are raised or lowered by the flat-space metric $\eta_{ab}$,
\begin{equation}
\label{flateta}
\eta_{ab}=E^\mu_a E^\nu_b g_{\mu\nu}=\begin{pmatrix}
                                       0 & -1 & 0 & 0 \\
                                       -1 & 0 & 0 & 0 \\
                                       0 & 0 & 0 & 1 \\
                                       0 & 0 & 1 & 0
                                     \end{pmatrix}.
\end{equation}
For a weak, plane and {\em null} wave  propagating along the $z$ direction,
the linearized Riemann tensor depends on $u=t-z$ only, so $R_{abcd,p}=0$ , where $(a,b,c,d)$ range over $(k,l,m,\bar{m})$, while $(p,q,r,\ldots)$ range over $(k,m,\bar{m})$ only.
To the linear approximation of $h_{\mu\nu}$, the covariant derivative in Bianchi identity becomes the coordinate derivative \cite{Eardley:1974nw},
\begin{equation}
\label{npeq4}
R_{ab(pq;l)}=R_{ab(pq,l)}=\frac{1}{3}(R_{abpq,l}+R_{abql,p}+R_{ablp,q})=\frac{1}{3}R_{abpq,l}=0,
\end{equation}
so $R_{abpq}$ is a constant. For propagating gravitational waves, we have $R_{abpq}=R_{pqab}=0$. Therefore, if no $l$ index appears in the first two indices
or the last two indices, then the Riemann tensor is 0, i.e., 
$R_{kk}=R_{km}=R_{k\bar{m}}=R_{mm}=R_{m\bar{m}}=R_{\bar{m}\bar{m}}=0$.
The nonzero components of the Riemann tensor are $R_{plql}$, i.e., $R_{plql}\neq 0$, so the six
degrees of freedom are $\Psi_4(u)$, $\Phi_{22}(u)$, $\Psi_3(u)$ and $\Psi_2(u)$. Note that
the formula \eqref{npeq4} and the results for the degrees of freedom are obtained
under the assumption that the waves are null waves. If there are massive degrees of freedom, then the the above derivation is
not applicable and we cannot use the above results to classify the propagating degrees of freedom.


\begin{thebibliography}{48}%
\makeatletter
\providecommand \@ifxundefined [1]{%
 \@ifx{#1\undefined}
}%
\providecommand \@ifnum [1]{%
 \ifnum #1\expandafter \@firstoftwo
 \else \expandafter \@secondoftwo
 \fi
}%
\providecommand \@ifx [1]{%
 \ifx #1\expandafter \@firstoftwo
 \else \expandafter \@secondoftwo
 \fi
}%
\providecommand \natexlab [1]{#1}%
\providecommand \enquote  [1]{``#1''}%
\providecommand \bibnamefont  [1]{#1}%
\providecommand \bibfnamefont [1]{#1}%
\providecommand \citenamefont [1]{#1}%
\providecommand \href@noop [0]{\@secondoftwo}%
\providecommand \href [0]{\begingroup \@sanitize@url \@href}%
\providecommand \@href[1]{\@@startlink{#1}\@@href}%
\providecommand \@@href[1]{\endgroup#1\@@endlink}%
\providecommand \@sanitize@url [0]{\catcode `\\12\catcode `\$12\catcode
  `\&12\catcode `\#12\catcode `\^12\catcode `\_12\catcode `\%12\relax}%
\providecommand \@@startlink[1]{}%
\providecommand \@@endlink[0]{}%
\providecommand \url  [0]{\begingroup\@sanitize@url \@url }%
\providecommand \@url [1]{\endgroup\@href {#1}{\urlprefix }}%
\providecommand \urlprefix  [0]{URL }%
\providecommand \Eprint [0]{\href }%
\providecommand \doibase [0]{http://dx.doi.org/}%
\providecommand \selectlanguage [0]{\@gobble}%
\providecommand \bibinfo  [0]{\@secondoftwo}%
\providecommand \bibfield  [0]{\@secondoftwo}%
\providecommand \translation [1]{[#1]}%
\providecommand \BibitemOpen [0]{}%
\providecommand \bibitemStop [0]{}%
\providecommand \bibitemNoStop [0]{.\EOS\space}%
\providecommand \EOS [0]{\spacefactor3000\relax}%
\providecommand \BibitemShut  [1]{\csname bibitem#1\endcsname}%
\let\auto@bib@innerbib\@empty
\bibitem [{\citenamefont {Abbott}\ \emph {et~al.}(2016)\citenamefont {Abbott}
  \emph {et~al.}}]{Abbott:2016blz}%
  \BibitemOpen
  \bibfield  {author} {\bibinfo {author} {\bibfnamefont {B.~P.}\ \bibnamefont
  {Abbott}} \emph {et~al.} (\bibinfo {collaboration} {LIGO Scientific and Virgo
  Collaborations}),\ }\href {\doibase 10.1103/PhysRevLett.116.061102}
  {\bibfield  {journal} {\bibinfo  {journal} {Phys. Rev. Lett.}\ }\textbf
  {\bibinfo {volume} {116}},\ \bibinfo {pages} {061102} (\bibinfo {year}
  {2016})},\ \Eprint {http://arxiv.org/abs/1602.03837} {arXiv:1602.03837
  [gr-qc]} \BibitemShut {NoStop}%
\bibitem [{\citenamefont {Newman}\ and\ \citenamefont
  {Penrose}(1962)}]{Newman:1961qr}%
  \BibitemOpen
  \bibfield  {author} {\bibinfo {author} {\bibfnamefont {E.}~\bibnamefont
  {Newman}}\ and\ \bibinfo {author} {\bibfnamefont {R.}~\bibnamefont
  {Penrose}},\ }\href {\doibase 10.1063/1.1724257} {\bibfield  {journal}
  {\bibinfo  {journal} {J. Math. Phys.}\ }\textbf {\bibinfo {volume} {3}},\
  \bibinfo {pages} {566} (\bibinfo {year} {1962})}\BibitemShut {NoStop}%
\bibitem [{\citenamefont {Eardley}\ \emph {et~al.}(1973)\citenamefont
  {Eardley}, \citenamefont {Lee},\ and\ \citenamefont
  {Lightman}}]{Eardley:1974nw}%
  \BibitemOpen
  \bibfield  {author} {\bibinfo {author} {\bibfnamefont {D.~M.}\ \bibnamefont
  {Eardley}}, \bibinfo {author} {\bibfnamefont {D.~L.}\ \bibnamefont {Lee}}, \
  and\ \bibinfo {author} {\bibfnamefont {A.~P.}\ \bibnamefont {Lightman}},\
  }\href {\doibase 10.1103/PhysRevD.8.3308} {\bibfield  {journal} {\bibinfo
  {journal} {Phys. Rev. D}\ }\textbf {\bibinfo {volume} {8}},\ \bibinfo {pages}
  {3308} (\bibinfo {year} {1973})}\BibitemShut {NoStop}%
\bibitem [{\citenamefont {Brans}\ and\ \citenamefont
  {Dicke}(1961)}]{Brans:1961sx}%
  \BibitemOpen
  \bibfield  {author} {\bibinfo {author} {\bibfnamefont {C.}~\bibnamefont
  {Brans}}\ and\ \bibinfo {author} {\bibfnamefont {R.}~\bibnamefont {Dicke}},\
  }\href {\doibase 10.1103/PhysRev.124.925} {\bibfield  {journal} {\bibinfo
  {journal} {Phys. Rev.}\ }\textbf {\bibinfo {volume} {124}},\ \bibinfo {pages}
  {925} (\bibinfo {year} {1961})}\BibitemShut {NoStop}%
\bibitem [{\citenamefont {Utiyama}\ and\ \citenamefont
  {DeWitt}(1962)}]{Utiyama:1962sn}%
  \BibitemOpen
  \bibfield  {author} {\bibinfo {author} {\bibfnamefont {R.}~\bibnamefont
  {Utiyama}}\ and\ \bibinfo {author} {\bibfnamefont {B.~S.}\ \bibnamefont
  {DeWitt}},\ }\href {\doibase 10.1063/1.1724264} {\bibfield  {journal}
  {\bibinfo  {journal} {J. Math. Phys.}\ }\textbf {\bibinfo {volume} {3}},\
  \bibinfo {pages} {608} (\bibinfo {year} {1962})}\BibitemShut {NoStop}%
\bibitem [{\citenamefont {Stelle}(1977)}]{Stelle:1976gc}%
  \BibitemOpen
  \bibfield  {author} {\bibinfo {author} {\bibfnamefont {K.}~\bibnamefont
  {Stelle}},\ }\href {\doibase 10.1103/PhysRevD.16.953} {\bibfield  {journal}
  {\bibinfo  {journal} {Phys. Rev. D}\ }\textbf {\bibinfo {volume} {16}},\
  \bibinfo {pages} {953} (\bibinfo {year} {1977})}\BibitemShut {NoStop}%
\bibitem [{\citenamefont {Stelle}(1978)}]{Stelle:1977ry}%
  \BibitemOpen
  \bibfield  {author} {\bibinfo {author} {\bibfnamefont {K.~S.}\ \bibnamefont
  {Stelle}},\ }\href {\doibase 10.1007/BF00760427} {\bibfield  {journal}
  {\bibinfo  {journal} {Gen. Relativ. Gravit.}\ }\textbf {\bibinfo {volume} {9}},\
  \bibinfo {pages} {353} (\bibinfo {year} {1978})}\BibitemShut {NoStop}%
\bibitem [{\citenamefont {O'~Hanlon}(1972)}]{OHanlon:1972xqa}%
  \BibitemOpen
  \bibfield  {author} {\bibinfo {author} {\bibfnamefont {J.}~\bibnamefont
  {O'~Hanlon}},\ }\href {\doibase 10.1103/PhysRevLett.29.137} {\bibfield
  {journal} {\bibinfo  {journal} {Phys. Rev. Lett.}\ }\textbf {\bibinfo
  {volume} {29}},\ \bibinfo {pages} {137} (\bibinfo {year} {1972})}\BibitemShut
  {NoStop}%
\bibitem [{\citenamefont {Teyssandier}\ and\ \citenamefont
  {Tourrenc}(1983)}]{Teyssandier:1983zz}%
  \BibitemOpen
  \bibfield  {author} {\bibinfo {author} {\bibfnamefont {P.}~\bibnamefont
  {Teyssandier}}\ and\ \bibinfo {author} {\bibfnamefont {P.}~\bibnamefont
  {Tourrenc}},\ }\href {\doibase 10.1063/1.525659} {\bibfield  {journal}
  {\bibinfo  {journal} {J. Math. Phys.}\ }\textbf {\bibinfo {volume} {24}},\
  \bibinfo {pages} {2793} (\bibinfo {year} {1983})}\BibitemShut {NoStop}%
\bibitem [{\citenamefont {Corda}(2007)}]{Corda:2007hi}%
  \BibitemOpen
  \bibfield  {author} {\bibinfo {author} {\bibfnamefont {C.}~\bibnamefont
  {Corda}},\ }\href {\doibase 10.1088/1475-7516/2007/04/009} {\bibfield
  {journal} {\bibinfo  {journal} {JCAP}\ }\textbf {\bibinfo {volume} {0704}},\
  \bibinfo {pages} {009} (\bibinfo {year} {2007})},\ \Eprint
  {http://arxiv.org/abs/astro-ph/0703644} {arXiv:astro-ph/0703644 [astro-ph]}
  \BibitemShut {NoStop}%
\bibitem [{\citenamefont {Corda}(2008)}]{Corda:2007nr}%
  \BibitemOpen
  \bibfield  {author} {\bibinfo {author} {\bibfnamefont {C.}~\bibnamefont
  {Corda}},\ }\href {\doibase 10.1142/S0217751X08038603} {\bibfield  {journal}
  {\bibinfo  {journal} {Int. J. Mod. Phys. A}\ }\textbf {\bibinfo {volume}
  {23}},\ \bibinfo {pages} {1521} (\bibinfo {year} {2008})},\ \Eprint
  {http://arxiv.org/abs/0711.4917} {arXiv:0711.4917 [gr-qc]} \BibitemShut
  {NoStop}%
\bibitem [{\citenamefont {Capozziello}\ \emph {et~al.}(2008)\citenamefont
  {Capozziello}, \citenamefont {Corda},\ and\ \citenamefont
  {De~Laurentis}}]{Capozziello:2008rq}%
  \BibitemOpen
  \bibfield  {author} {\bibinfo {author} {\bibfnamefont {S.}~\bibnamefont
  {Capozziello}}, \bibinfo {author} {\bibfnamefont {C.}~\bibnamefont {Corda}},
  \ and\ \bibinfo {author} {\bibfnamefont {M.~F.}\ \bibnamefont
  {De~Laurentis}},\ }\href {\doibase 10.1016/j.physletb.2008.10.001} {\bibfield
   {journal} {\bibinfo  {journal} {Phys. Lett. B}\ }\textbf {\bibinfo {volume}
  {669}},\ \bibinfo {pages} {255} (\bibinfo {year} {2008})},\ \Eprint
  {http://arxiv.org/abs/0812.2272} {arXiv:0812.2272 [astro-ph]} \BibitemShut
  {NoStop}%
\bibitem [{\citenamefont {Capozziello}\ \emph {et~al.}(2010)\citenamefont
  {Capozziello}, \citenamefont {Cianci}, \citenamefont {De~Laurentis},\ and\
  \citenamefont {Vignolo}}]{Capozziello:2010iy}%
  \BibitemOpen
  \bibfield  {author} {\bibinfo {author} {\bibfnamefont {S.}~\bibnamefont
  {Capozziello}}, \bibinfo {author} {\bibfnamefont {R.}~\bibnamefont {Cianci}},
  \bibinfo {author} {\bibfnamefont {M.}~\bibnamefont {De~Laurentis}}, \ and\
  \bibinfo {author} {\bibfnamefont {S.}~\bibnamefont {Vignolo}},\ }\href
  {\doibase 10.1140/epjc/s10052-010-1412-5} {\bibfield  {journal} {\bibinfo
  {journal} {Eur. Phys. J. C}\ }\textbf {\bibinfo {volume} {70}},\ \bibinfo
  {pages} {341} (\bibinfo {year} {2010})},\ \Eprint
  {http://arxiv.org/abs/1007.3670} {arXiv:1007.3670 [gr-qc]} \BibitemShut
  {NoStop}%
\bibitem [{\citenamefont {Prasia}\ and\ \citenamefont
  {Kuriakose}(2014)}]{Prasia:2014zya}%
  \BibitemOpen
  \bibfield  {author} {\bibinfo {author} {\bibfnamefont {P.}~\bibnamefont
  {Prasia}}\ and\ \bibinfo {author} {\bibfnamefont {V.~C.}\ \bibnamefont
  {Kuriakose}},\ }\href {\doibase 10.1142/S0218271814500370} {\bibfield
  {journal} {\bibinfo  {journal} {Int. J. Mod. Phys. D}\ }\textbf {\bibinfo
  {volume} {23}},\ \bibinfo {pages} {1450037} (\bibinfo {year}
  {2014})}\BibitemShut {NoStop}%
\bibitem [{\citenamefont {Alves}\ \emph {et~al.}(2009)\citenamefont {Alves},
  \citenamefont {Miranda},\ and\ \citenamefont {de~Araujo}}]{Alves:2009eg}%
  \BibitemOpen
  \bibfield  {author} {\bibinfo {author} {\bibfnamefont {M.~E.~S.}\
  \bibnamefont {Alves}}, \bibinfo {author} {\bibfnamefont {O.~D.}\ \bibnamefont
  {Miranda}}, \ and\ \bibinfo {author} {\bibfnamefont {J.~C.~N.}\ \bibnamefont
  {de~Araujo}},\ }\href {\doibase 10.1016/j.physletb.2009.08.005} {\bibfield
  {journal} {\bibinfo  {journal} {Phys. Lett. B}\ }\textbf {\bibinfo {volume}
  {679}},\ \bibinfo {pages} {401} (\bibinfo {year} {2009})},\ \Eprint
  {http://arxiv.org/abs/0908.0861} {arXiv:0908.0861 [gr-qc]} \BibitemShut
  {NoStop}%
\bibitem [{\citenamefont {Alves}\ \emph {et~al.}(2010)\citenamefont {Alves},
  \citenamefont {Miranda},\ and\ \citenamefont {de~Araujo}}]{Alves:2010ms}%
  \BibitemOpen
  \bibfield  {author} {\bibinfo {author} {\bibfnamefont {M.~E.~S.}\
  \bibnamefont {Alves}}, \bibinfo {author} {\bibfnamefont {O.~D.}\ \bibnamefont
  {Miranda}}, \ and\ \bibinfo {author} {\bibfnamefont {J.~C.~N.}\ \bibnamefont
  {de~Araujo}},\ }\href {\doibase 10.1088/0264-9381/27/14/145010} {\bibfield
  {journal} {\bibinfo  {journal} {Class. Quant. Grav.}\ }\textbf {\bibinfo
  {volume} {27}},\ \bibinfo {pages} {145010} (\bibinfo {year} {2010})},\
  \Eprint {http://arxiv.org/abs/1004.5580} {arXiv:1004.5580 [gr-qc]}
  \BibitemShut {NoStop}%
\bibitem [{\citenamefont {Kausar}\ \emph {et~al.}(2016)\citenamefont {Kausar},
  \citenamefont {Philippoz},\ and\ \citenamefont {Jetzer}}]{Rizwana:2016qdq}%
  \BibitemOpen
  \bibfield  {author} {\bibinfo {author} {\bibfnamefont {H.~R.}\ \bibnamefont
  {Kausar}}, \bibinfo {author} {\bibfnamefont {L.}~\bibnamefont {Philippoz}}, \
  and\ \bibinfo {author} {\bibfnamefont {P.}~\bibnamefont {Jetzer}},\ }\href
  {\doibase 10.1103/PhysRevD.93.124071} {\bibfield  {journal} {\bibinfo
  {journal} {Phys. Rev. D}\ }\textbf {\bibinfo {volume} {93}},\ \bibinfo
  {pages} {124071} (\bibinfo {year} {2016})},\ \Eprint
  {http://arxiv.org/abs/1606.07000} {arXiv:1606.07000 [gr-qc]} \BibitemShut
  {NoStop}%
\bibitem [{\citenamefont {Myung}(2016)}]{Myung:2016zdl}%
  \BibitemOpen
  \bibfield  {author} {\bibinfo {author} {\bibfnamefont {Y.~S.}\ \bibnamefont
  {Myung}},\ }\href {\doibase 10.1155/2016/3901734} {\bibfield  {journal}
  {\bibinfo  {journal} {Adv. High Energy Phys.}\ }\textbf {\bibinfo {volume}
  {2016}},\ \bibinfo {pages} {3901734} (\bibinfo {year} {2016})},\ \Eprint
  {http://arxiv.org/abs/1608.01764} {arXiv:1608.01764 [gr-qc]} \BibitemShut
  {NoStop}%
\bibitem [{\citenamefont {Starobinsky}(1980)}]{starobinskyfr}%
  \BibitemOpen
  \bibfield  {author} {\bibinfo {author} {\bibfnamefont {A.~A.}\ \bibnamefont
  {Starobinsky}},\ }\href {\doibase 10.1016/0370-2693(80)90670-X} {\bibfield
  {journal} {\bibinfo  {journal} {Phys. Lett. B.}\ }\textbf {\bibinfo {volume}
  {91}},\ \bibinfo {pages} {99} (\bibinfo {year} {1980})}\BibitemShut {NoStop}%
\bibitem [{\citenamefont {Ade}\ \emph {et~al.}(2016)\citenamefont {Ade} \emph
  {et~al.}}]{Ade:2015lrj}%
  \BibitemOpen
  \bibfield  {author} {\bibinfo {author} {\bibfnamefont {P.~A.~R.}\
  \bibnamefont {Ade}} \emph {et~al.} (\bibinfo {collaboration} {Planck}),\
  }\href {\doibase 10.1051/0004-6361/201525898} {\bibfield  {journal} {\bibinfo
   {journal} {Astron. Astrophys.}\ }\textbf {\bibinfo {volume} {594}},\
  \bibinfo {pages} {A20} (\bibinfo {year} {2016})},\ \Eprint
  {http://arxiv.org/abs/1502.02114} {arXiv:1502.02114 [astro-ph.CO]}
  \BibitemShut {NoStop}%
\bibitem [{\citenamefont {Riess}\ \emph {et~al.}(1998)\citenamefont {Riess}
  \emph {et~al.}}]{hzsst98}%
  \BibitemOpen
  \bibfield  {author} {\bibinfo {author} {\bibfnamefont {A.~G.}\ \bibnamefont
  {Riess}} \emph {et~al.} (\bibinfo {collaboration} {Supernova Search Team}),\
  }\href {\doibase 10.1086/300499} {\bibfield  {journal} {\bibinfo  {journal}
  {Astron. J.}\ }\textbf {\bibinfo {volume} {116}},\ \bibinfo {pages} {1009}
  (\bibinfo {year} {1998})},\ \Eprint {http://arxiv.org/abs/astro-ph/9805201}
  {arXiv:astro-ph/9805201 [astro-ph]} \BibitemShut {NoStop}%
\bibitem [{\citenamefont {Perlmutter}\ \emph {et~al.}(1999)\citenamefont
  {Perlmutter} \emph {et~al.}}]{scpsn98}%
  \BibitemOpen
  \bibfield  {author} {\bibinfo {author} {\bibfnamefont {S.}~\bibnamefont
  {Perlmutter}} \emph {et~al.} (\bibinfo {collaboration} {Supernova Cosmology
  Project}),\ }\href {\doibase 10.1086/307221} {\bibfield  {journal} {\bibinfo
  {journal} {Astrophys. J.}\ }\textbf {\bibinfo {volume} {517}},\ \bibinfo
  {pages} {565} (\bibinfo {year} {1999})},\ \Eprint
  {http://arxiv.org/abs/astro-ph/9812133} {arXiv:astro-ph/9812133 [astro-ph]}
  \BibitemShut {NoStop}%
\bibitem [{\citenamefont {Carroll}\ \emph {et~al.}(2004)\citenamefont
  {Carroll}, \citenamefont {Duvvuri}, \citenamefont {Trodden},\ and\
  \citenamefont {Turner}}]{Carroll:2003wy}%
  \BibitemOpen
  \bibfield  {author} {\bibinfo {author} {\bibfnamefont {S.~M.}\ \bibnamefont
  {Carroll}}, \bibinfo {author} {\bibfnamefont {V.}~\bibnamefont {Duvvuri}},
  \bibinfo {author} {\bibfnamefont {M.}~\bibnamefont {Trodden}}, \ and\
  \bibinfo {author} {\bibfnamefont {M.~S.}\ \bibnamefont {Turner}},\ }\href
  {\doibase 10.1103/PhysRevD.70.043528} {\bibfield  {journal} {\bibinfo
  {journal} {Phys. Rev. D}\ }\textbf {\bibinfo {volume} {70}},\ \bibinfo
  {pages} {043528} (\bibinfo {year} {2004})},\ \Eprint
  {http://arxiv.org/abs/astro-ph/0306438} {arXiv:astro-ph/0306438 [astro-ph]}
  \BibitemShut {NoStop}%
\bibitem [{\citenamefont {Vollick}(2003)}]{Vollick:2003aw}%
  \BibitemOpen
  \bibfield  {author} {\bibinfo {author} {\bibfnamefont {D.~N.}\ \bibnamefont
  {Vollick}},\ }\href {\doibase 10.1103/PhysRevD.68.063510} {\bibfield
  {journal} {\bibinfo  {journal} {Phys. Rev. D}\ }\textbf {\bibinfo {volume}
  {68}},\ \bibinfo {pages} {063510} (\bibinfo {year} {2003})},\ \Eprint
  {http://arxiv.org/abs/astro-ph/0306630} {arXiv:astro-ph/0306630 [astro-ph]}
  \BibitemShut {NoStop}%
\bibitem [{\citenamefont {Flanagan}(2004)}]{Flanagan:2003rb}%
  \BibitemOpen
  \bibfield  {author} {\bibinfo {author} {\bibfnamefont {E.~E.}\ \bibnamefont
  {Flanagan}},\ }\href {\doibase 10.1103/PhysRevLett.92.071101} {\bibfield
  {journal} {\bibinfo  {journal} {Phys. Rev. Lett.}\ }\textbf {\bibinfo
  {volume} {92}},\ \bibinfo {pages} {071101} (\bibinfo {year} {2004})},\
  \Eprint {http://arxiv.org/abs/astro-ph/0308111} {arXiv:astro-ph/0308111
  [astro-ph]} \BibitemShut {NoStop}%
\bibitem [{\citenamefont {Nojiri}\ and\ \citenamefont
  {Odintsov}(2003)}]{Nojiri:2003ft}%
  \BibitemOpen
  \bibfield  {author} {\bibinfo {author} {\bibfnamefont {S.}~\bibnamefont
  {Nojiri}}\ and\ \bibinfo {author} {\bibfnamefont {S.~D.}\ \bibnamefont
  {Odintsov}},\ }\href {\doibase 10.1103/PhysRevD.68.123512} {\bibfield
  {journal} {\bibinfo  {journal} {Phys. Rev. D}\ }\textbf {\bibinfo {volume}
  {68}},\ \bibinfo {pages} {123512} (\bibinfo {year} {2003})},\ \Eprint
  {http://arxiv.org/abs/hep-th/0307288} {arXiv:hep-th/0307288 [hep-th]}
  \BibitemShut {NoStop}%
\bibitem [{\citenamefont {Chiba}(2003)}]{Chiba:2003ir}%
  \BibitemOpen
  \bibfield  {author} {\bibinfo {author} {\bibfnamefont {T.}~\bibnamefont
  {Chiba}},\ }\href {\doibase 10.1016/j.physletb.2003.09.033} {\bibfield
  {journal} {\bibinfo  {journal} {Phys. Lett. B}\ }\textbf {\bibinfo {volume}
  {575}},\ \bibinfo {pages} {1} (\bibinfo {year} {2003})},\ \Eprint
  {http://arxiv.org/abs/astro-ph/0307338} {arXiv:astro-ph/0307338 [astro-ph]}
  \BibitemShut {NoStop}%
\bibitem [{\citenamefont {Erickcek}\ \emph {et~al.}(2006)\citenamefont
  {Erickcek}, \citenamefont {Smith},\ and\ \citenamefont
  {Kamionkowski}}]{Erickcek:2006vf}%
  \BibitemOpen
  \bibfield  {author} {\bibinfo {author} {\bibfnamefont {A.~L.}\ \bibnamefont
  {Erickcek}}, \bibinfo {author} {\bibfnamefont {T.~L.}\ \bibnamefont {Smith}},
  \ and\ \bibinfo {author} {\bibfnamefont {M.}~\bibnamefont {Kamionkowski}},\
  }\href {\doibase 10.1103/PhysRevD.74.121501} {\bibfield  {journal} {\bibinfo
  {journal} {Phys. Rev. D}\ }\textbf {\bibinfo {volume} {74}},\ \bibinfo
  {pages} {121501} (\bibinfo {year} {2006})},\ \Eprint
  {http://arxiv.org/abs/astro-ph/0610483} {arXiv:astro-ph/0610483 [astro-ph]}
  \BibitemShut {NoStop}%
\bibitem [{\citenamefont {Hu}\ and\ \citenamefont {Sawicki}(2007)}]{Hu:2007nk}%
  \BibitemOpen
  \bibfield  {author} {\bibinfo {author} {\bibfnamefont {W.}~\bibnamefont
  {Hu}}\ and\ \bibinfo {author} {\bibfnamefont {I.}~\bibnamefont {Sawicki}},\
  }\href {\doibase 10.1103/PhysRevD.76.064004} {\bibfield  {journal} {\bibinfo
  {journal} {Phys. Rev. D}\ }\textbf {\bibinfo {volume} {76}},\ \bibinfo
  {pages} {064004} (\bibinfo {year} {2007})},\ \Eprint
  {http://arxiv.org/abs/0705.1158} {arXiv:0705.1158 [astro-ph]} \BibitemShut
  {NoStop}%
\bibitem [{\citenamefont {Starobinsky}(2007)}]{Starobinsky:2007hu}%
  \BibitemOpen
  \bibfield  {author} {\bibinfo {author} {\bibfnamefont {A.~A.}\ \bibnamefont
  {Starobinsky}},\ }\href {\doibase 10.1134/S0021364007150027} {\bibfield
  {journal} {\bibinfo  {journal} {JETP Lett.}\ }\textbf {\bibinfo {volume}
  {86}},\ \bibinfo {pages} {157} (\bibinfo {year} {2007})},\ \Eprint
  {http://arxiv.org/abs/0706.2041} {arXiv:0706.2041 [astro-ph]} \BibitemShut
  {NoStop}%
\bibitem [{\citenamefont {Capozziello}\ \emph {et~al.}(2009)\citenamefont
  {Capozziello}, \citenamefont {De~Laurentis}, \citenamefont {Nojiri},\ and\
  \citenamefont {Odintsov}}]{Capozziello:2008fn}%
  \BibitemOpen
  \bibfield  {author} {\bibinfo {author} {\bibfnamefont {S.}~\bibnamefont
  {Capozziello}}, \bibinfo {author} {\bibfnamefont {M.}~\bibnamefont
  {De~Laurentis}}, \bibinfo {author} {\bibfnamefont {S.}~\bibnamefont
  {Nojiri}}, \ and\ \bibinfo {author} {\bibfnamefont {S.~D.}\ \bibnamefont
  {Odintsov}},\ }\href {\doibase 10.1007/s10714-009-0758-1} {\bibfield
  {journal} {\bibinfo  {journal} {Gen. Relativ. Gravit.}\ }\textbf {\bibinfo {volume}
  {41}},\ \bibinfo {pages} {2313} (\bibinfo {year} {2009})},\ \Eprint
  {http://arxiv.org/abs/0808.1335} {arXiv:0808.1335 [hep-th]} \BibitemShut
  {NoStop}%
\bibitem [{\citenamefont {Cognola}\ \emph {et~al.}(2008)\citenamefont
  {Cognola}, \citenamefont {Elizalde}, \citenamefont {Nojiri}, \citenamefont
  {Odintsov}, \citenamefont {Sebastiani},\ and\ \citenamefont
  {Zerbini}}]{Cognola:2007zu}%
  \BibitemOpen
  \bibfield  {author} {\bibinfo {author} {\bibfnamefont {G.}~\bibnamefont
  {Cognola}}, \bibinfo {author} {\bibfnamefont {E.}~\bibnamefont {Elizalde}},
  \bibinfo {author} {\bibfnamefont {S.}~\bibnamefont {Nojiri}}, \bibinfo
  {author} {\bibfnamefont {S.~D.}\ \bibnamefont {Odintsov}}, \bibinfo {author}
  {\bibfnamefont {L.}~\bibnamefont {Sebastiani}}, \ and\ \bibinfo {author}
  {\bibfnamefont {S.}~\bibnamefont {Zerbini}},\ }\href {\doibase
  10.1103/PhysRevD.77.046009} {\bibfield  {journal} {\bibinfo  {journal} {Phys.
  Rev. D}\ }\textbf {\bibinfo {volume} {77}},\ \bibinfo {pages} {046009}
  (\bibinfo {year} {2008})},\ \Eprint {http://arxiv.org/abs/0712.4017}
  {arXiv:0712.4017 [hep-th]} \BibitemShut {NoStop}%
\bibitem [{\citenamefont {Nojiri}\ and\ \citenamefont
  {Odintsov}(2008)}]{Nojiri:2007cq}%
  \BibitemOpen
  \bibfield  {author} {\bibinfo {author} {\bibfnamefont {S.}~\bibnamefont
  {Nojiri}}\ and\ \bibinfo {author} {\bibfnamefont {S.~D.}\ \bibnamefont
  {Odintsov}},\ }\href {\doibase 10.1103/PhysRevD.77.026007} {\bibfield
  {journal} {\bibinfo  {journal} {Phys. Rev. D}\ }\textbf {\bibinfo {volume}
  {77}},\ \bibinfo {pages} {026007} (\bibinfo {year} {2008})},\ \Eprint
  {http://arxiv.org/abs/0710.1738} {arXiv:0710.1738 [hep-th]} \BibitemShut
  {NoStop}%
\bibitem [{\citenamefont {Myrzakulov}\ \emph {et~al.}(2015)\citenamefont
  {Myrzakulov}, \citenamefont {Sebastiani},\ and\ \citenamefont
  {Vagnozzi}}]{Myrzakulov:2015qaa}%
  \BibitemOpen
  \bibfield  {author} {\bibinfo {author} {\bibfnamefont {R.}~\bibnamefont
  {Myrzakulov}}, \bibinfo {author} {\bibfnamefont {L.}~\bibnamefont
  {Sebastiani}}, \ and\ \bibinfo {author} {\bibfnamefont {S.}~\bibnamefont
  {Vagnozzi}},\ }\href {\doibase 10.1140/epjc/s10052-015-3672-6} {\bibfield
  {journal} {\bibinfo  {journal} {Eur. Phys. J. C}\ }\textbf {\bibinfo {volume}
  {75}},\ \bibinfo {pages} {444} (\bibinfo {year} {2015})},\ \Eprint
  {http://arxiv.org/abs/1504.07984} {arXiv:1504.07984 [gr-qc]} \BibitemShut
  {NoStop}%
\bibitem [{\citenamefont {Yi}\ and\ \citenamefont {Gong}(2016)}]{Yi:2016jqr}%
  \BibitemOpen
  \bibfield  {author} {\bibinfo {author} {\bibfnamefont {Z.}~\bibnamefont
  {Yi}}\ and\ \bibinfo {author} {\bibfnamefont {Y.}~\bibnamefont {Gong}},\
  }\href {\doibase 10.1103/PhysRevD.94.103527} {\bibfield  {journal} {\bibinfo
  {journal} {Phys. Rev. D}\ }\textbf {\bibinfo {volume} {94}},\ \bibinfo
  {pages} {103527} (\bibinfo {year} {2016})},\ \Eprint
  {http://arxiv.org/abs/1608.05922} {arXiv:1608.05922 [gr-qc]} \BibitemShut
  {NoStop}%
\bibitem [{\citenamefont {Goldhaber}\ and\ \citenamefont
  {Nieto}(1974)}]{Goldhaber:1974wg}%
  \BibitemOpen
  \bibfield  {author} {\bibinfo {author} {\bibfnamefont {A.~S.}\ \bibnamefont
  {Goldhaber}}\ and\ \bibinfo {author} {\bibfnamefont {M.~M.}\ \bibnamefont
  {Nieto}},\ }\href {\doibase 10.1103/PhysRevD.9.1119} {\bibfield  {journal}
  {\bibinfo  {journal} {Phys. Rev. D}\ }\textbf {\bibinfo {volume} {9}},\
  \bibinfo {pages} {1119} (\bibinfo {year} {1974})}\BibitemShut {NoStop}%
\bibitem [{\citenamefont {Berry}\ and\ \citenamefont
  {Gair}(2011)}]{Berry:2011pb}%
  \BibitemOpen
  \bibfield  {author} {\bibinfo {author} {\bibfnamefont {C.~P.~L.}\
  \bibnamefont {Berry}}\ and\ \bibinfo {author} {\bibfnamefont {J.~R.}\
  \bibnamefont {Gair}},\ }\href {\doibase 10.1103/PhysRevD.85.089906,
  10.1103/PhysRevD.83.104022} {\bibfield  {journal} {\bibinfo  {journal} {Phys.
  Rev. D}\ }\textbf {\bibinfo {volume} {83}},\ \bibinfo {pages} {104022}
  (\bibinfo {year} {2011})},\ \bibinfo {note} {[Erratum: Phys. Rev. D 85,
  089906 (2012)]},\ \Eprint {http://arxiv.org/abs/1104.0819} {arXiv:1104.0819
  [gr-qc]} \BibitemShut {NoStop}%
\bibitem [{\citenamefont {Rivers}(1964)}]{spin2}%
  \BibitemOpen
  \bibfield  {author} {\bibinfo {author} {\bibfnamefont {R.}~\bibnamefont
  {Rivers}},\ }\href@noop {} {\bibfield  {journal} {\bibinfo  {journal} {IL
  Nuovo Cimento}\ }\textbf {\bibinfo {volume} {34}},\ \bibinfo {pages} {386}
  (\bibinfo {year} {1964})}\BibitemShut {NoStop}%
\bibitem [{\citenamefont {Alonso}\ \emph {et~al.}(1994)\citenamefont {Alonso},
  \citenamefont {Barbero~G.}, \citenamefont {Julve},\ and\ \citenamefont
  {Tiemblo}}]{Alonso:1994tr}%
  \BibitemOpen
  \bibfield  {author} {\bibinfo {author} {\bibfnamefont {J.}~\bibnamefont
  {Alonso}}, \bibinfo {author} {\bibfnamefont {J.~F.}\ \bibnamefont
  {Barbero~G.}}, \bibinfo {author} {\bibfnamefont {J.}~\bibnamefont {Julve}}, \
  and\ \bibinfo {author} {\bibfnamefont {A.}~\bibnamefont {Tiemblo}},\ }\href
  {\doibase 10.1088/0264-9381/11/4/007} {\bibfield  {journal} {\bibinfo
  {journal} {Class. Quant. Grav.}\ }\textbf {\bibinfo {volume} {11}},\ \bibinfo
  {pages} {865} (\bibinfo {year} {1994})}\BibitemShut {NoStop}%
\bibitem [{\citenamefont {Ezawa}\ \emph {et~al.}(1999)\citenamefont {Ezawa},
  \citenamefont {Kajihara}, \citenamefont {Kiminami}, \citenamefont {Soda},\
  and\ \citenamefont {Yano}}]{Ezawa:1998ax}%
  \BibitemOpen
  \bibfield  {author} {\bibinfo {author} {\bibfnamefont {Y.}~\bibnamefont
  {Ezawa}}, \bibinfo {author} {\bibfnamefont {M.}~\bibnamefont {Kajihara}},
  \bibinfo {author} {\bibfnamefont {M.}~\bibnamefont {Kiminami}}, \bibinfo
  {author} {\bibfnamefont {J.}~\bibnamefont {Soda}}, \ and\ \bibinfo {author}
  {\bibfnamefont {T.}~\bibnamefont {Yano}},\ }\href {\doibase
  10.1088/0264-9381/16/4/003} {\bibfield  {journal} {\bibinfo  {journal}
  {Class. Quant. Grav.}\ }\textbf {\bibinfo {volume} {16}},\ \bibinfo {pages}
  {1127} (\bibinfo {year} {1999})},\ \Eprint
  {http://arxiv.org/abs/gr-qc/9801084} {arXiv:gr-qc/9801084 [gr-qc]}
  \BibitemShut {NoStop}%
\bibitem [{\citenamefont {Ezawa}\ \emph {et~al.}(2006)\citenamefont {Ezawa},
  \citenamefont {Iwasaki}, \citenamefont {Ohkuwa}, \citenamefont {Watanabe},
  \citenamefont {Yamada},\ and\ \citenamefont {Yano}}]{Ezawa:2005zr}%
  \BibitemOpen
  \bibfield  {author} {\bibinfo {author} {\bibfnamefont {Y.}~\bibnamefont
  {Ezawa}}, \bibinfo {author} {\bibfnamefont {H.}~\bibnamefont {Iwasaki}},
  \bibinfo {author} {\bibfnamefont {Y.}~\bibnamefont {Ohkuwa}}, \bibinfo
  {author} {\bibfnamefont {S.}~\bibnamefont {Watanabe}}, \bibinfo {author}
  {\bibfnamefont {N.}~\bibnamefont {Yamada}}, \ and\ \bibinfo {author}
  {\bibfnamefont {T.}~\bibnamefont {Yano}},\ }\href {\doibase
  10.1088/0264-9381/23/9/028} {\bibfield  {journal} {\bibinfo  {journal}
  {Class. Quant. Grav.}\ }\textbf {\bibinfo {volume} {23}},\ \bibinfo {pages}
  {3205} (\bibinfo {year} {2006})},\ \Eprint
  {http://arxiv.org/abs/gr-qc/0507060} {arXiv:gr-qc/0507060 [gr-qc]}
  \BibitemShut {NoStop}%
\bibitem [{\citenamefont {Ohkuwa}\ and\ \citenamefont
  {Ezawa}(2015)}]{Ohkuwa:2014mwa}%
  \BibitemOpen
  \bibfield  {author} {\bibinfo {author} {\bibfnamefont {Y.}~\bibnamefont
  {Ohkuwa}}\ and\ \bibinfo {author} {\bibfnamefont {Y.}~\bibnamefont {Ezawa}},\
  }\href {\doibase 10.1140/epjp/i2015-15077-5} {\bibfield  {journal} {\bibinfo
  {journal} {Eur. Phys. J. Plus}\ }\textbf {\bibinfo {volume} {130}},\ \bibinfo
  {pages} {77} (\bibinfo {year} {2015})},\ \Eprint
  {http://arxiv.org/abs/1412.4475} {arXiv:1412.4475 [gr-qc]} \BibitemShut
  {NoStop}%
\bibitem [{\citenamefont {Olmo}\ and\ \citenamefont
  {Sanchis-Alepuz}(2011)}]{Olmo:2011fh}%
  \BibitemOpen
  \bibfield  {author} {\bibinfo {author} {\bibfnamefont {G.~J.}\ \bibnamefont
  {Olmo}}\ and\ \bibinfo {author} {\bibfnamefont {H.}~\bibnamefont
  {Sanchis-Alepuz}},\ }\href {\doibase 10.1103/PhysRevD.83.104036} {\bibfield
  {journal} {\bibinfo  {journal} {Phys. Rev. D}\ }\textbf {\bibinfo {volume}
  {83}},\ \bibinfo {pages} {104036} (\bibinfo {year} {2011})},\ \Eprint
  {http://arxiv.org/abs/1101.3403} {arXiv:1101.3403 [gr-qc]} \BibitemShut
  {NoStop}%
\bibitem [{\citenamefont {Deruelle}\ \emph {et~al.}(2009)\citenamefont
  {Deruelle}, \citenamefont {Sendouda},\ and\ \citenamefont
  {Youssef}}]{Deruelle:2009pu}%
  \BibitemOpen
  \bibfield  {author} {\bibinfo {author} {\bibfnamefont {N.}~\bibnamefont
  {Deruelle}}, \bibinfo {author} {\bibfnamefont {Y.}~\bibnamefont {Sendouda}},
  \ and\ \bibinfo {author} {\bibfnamefont {A.}~\bibnamefont {Youssef}},\ }\href
  {\doibase 10.1103/PhysRevD.80.084032} {\bibfield  {journal} {\bibinfo
  {journal} {Phys. Rev. D}\ }\textbf {\bibinfo {volume} {80}},\ \bibinfo
  {pages} {084032} (\bibinfo {year} {2009})},\ \Eprint
  {http://arxiv.org/abs/0906.4983} {arXiv:0906.4983 [gr-qc]} \BibitemShut
  {NoStop}%
\bibitem [{\citenamefont {Deruelle}\ \emph {et~al.}(2010)\citenamefont
  {Deruelle}, \citenamefont {Sasaki}, \citenamefont {Sendouda},\ and\
  \citenamefont {Yamauchi}}]{Deruelle:2009zk}%
  \BibitemOpen
  \bibfield  {author} {\bibinfo {author} {\bibfnamefont {N.}~\bibnamefont
  {Deruelle}}, \bibinfo {author} {\bibfnamefont {M.}~\bibnamefont {Sasaki}},
  \bibinfo {author} {\bibfnamefont {Y.}~\bibnamefont {Sendouda}}, \ and\
  \bibinfo {author} {\bibfnamefont {D.}~\bibnamefont {Yamauchi}},\ }\href
  {\doibase 10.1143/PTP.123.169} {\bibfield  {journal} {\bibinfo  {journal}
  {Prog. Theor. Phys.}\ }\textbf {\bibinfo {volume} {123}},\ \bibinfo {pages}
  {169} (\bibinfo {year} {2010})},\ \Eprint {http://arxiv.org/abs/0908.0679}
  {arXiv:0908.0679 [hep-th]} \BibitemShut {NoStop}%
\bibitem [{\citenamefont {Sendouda}\ \emph {et~al.}(2011)\citenamefont
  {Sendouda}, \citenamefont {Deruelle}, \citenamefont {Sasaki},\ and\
  \citenamefont {Yamauchi}}]{Sendouda:2011hq}%
  \BibitemOpen
  \bibfield  {author} {\bibinfo {author} {\bibfnamefont {Y.}~\bibnamefont
  {Sendouda}}, \bibinfo {author} {\bibfnamefont {N.}~\bibnamefont {Deruelle}},
  \bibinfo {author} {\bibfnamefont {M.}~\bibnamefont {Sasaki}}, \ and\ \bibinfo
  {author} {\bibfnamefont {D.}~\bibnamefont {Yamauchi}},\ }
  \href {\doibase 10.1142/S2010194511000432}
  {\bibfield  {journal} {\bibinfo  {journal} {Int. J. Mod. Phys. Conf. Ser.}\
  }\textbf {\bibinfo {volume} {01}},\ \bibinfo {pages} {297} (\bibinfo {year}
  {2011})}\BibitemShut {NoStop}%
\bibitem [{\citenamefont {Arnowitt}\ \emph {et~al.}(1962)\citenamefont
  {Arnowitt}, \citenamefont {Deser},\ and\ \citenamefont {Misner}}]{adm1}%
  \BibitemOpen
  \bibfield  {author} {\bibinfo {author} {\bibfnamefont {R.}~\bibnamefont
  {Arnowitt}}, \bibinfo {author} {\bibfnamefont {S.}~\bibnamefont {Deser}}, \
  and\ \bibinfo {author} {\bibfnamefont {C.}~\bibnamefont {Misner}},\ }in\
  \href@noop {} {\emph {\bibinfo {booktitle} {Gravitation: An Introduction to
  Current Research}}},\ \bibinfo {editor} {edited by\ \bibinfo {editor}
  {\bibfnamefont {L.}~\bibnamefont {Witten}}}\ (\bibinfo  {publisher} {John
  Wiley and Sons, LTD},\ \bibinfo {address} {New York},\ \bibinfo {year}
  {1962})\ pp.\ \bibinfo {pages} {227--265}\BibitemShut {NoStop}%
\bibitem [{\citenamefont {Arnowitt}\ \emph {et~al.}(2008)\citenamefont
  {Arnowitt}, \citenamefont {Deser},\ and\ \citenamefont {Misner}}]{adm2}%
  \BibitemOpen
  \bibfield  {author} {\bibinfo {author} {\bibfnamefont {R.~L.}\ \bibnamefont
  {Arnowitt}}, \bibinfo {author} {\bibfnamefont {S.}~\bibnamefont {Deser}}, \
  and\ \bibinfo {author} {\bibfnamefont {C.~W.}\ \bibnamefont {Misner}},\
  }\href {\doibase 10.1007/s10714-008-0661-1} {\bibfield  {journal} {\bibinfo
  {journal} {Gen. Relativ. Gravit.}\ }\textbf {\bibinfo {volume} {40}},\ \bibinfo
  {pages} {1997} (\bibinfo {year} {2008})},\ \Eprint
  {http://arxiv.org/abs/gr-qc/0405109} {arXiv:gr-qc/0405109 [gr-qc]}
  \BibitemShut {NoStop}%
\end{thebibliography}

%

\end{document}